\documentclass[aps,pre,twocolumn,superscriptaddress,floatfix,showpacs]{revtex4}
\usepackage{graphicx}
\usepackage{bm}
\bibstyle{apsrev.bib}

\begin{document}
\title{Exons, introns and DNA thermodynamics}
\author{Enrico Carlon}
\affiliation{Interdisciplinary Research Institute c/o IEMN, Cit\'e
Scientifique BP 69, F-59652 Villeneuve d'Ascq, France}
\author{Mehdi Lejard Malki}
\affiliation{Interdisciplinary Research Institute c/o IEMN, Cit\'e
Scientifique BP 69, F-59652 Villeneuve d'Ascq, France}
\affiliation{Ecole Nationale Sup\'erieure de Physique de Strasbourg (ENSPS), 
Parc d'innovation, BP 10413, F-67412 Illkirch, France}
\author{Ralf Blossey}
\affiliation{Interdisciplinary Research Institute c/o IEMN, Cit\'e
Scientifique BP 69, F-59652 Villeneuve d'Ascq, France}

\date{\today}

\begin{abstract}
The genes of eukaryotes are characterized by protein coding fragments,
the exons, interrupted by introns, i.e. stretches of DNA which do not
carry useful information for protein synthesis.  We have analyzed
the melting behavior of randomly selected human cDNA sequences obtained
from genomic DNA by removing all introns. A clear correspondence is
observed between exons and melting domains. This finding may provide new
insights into the physical mechanisms underlying the evolution of genes.
\end{abstract}

\pacs{87.15.-v, 87.14.Gg, 05.70.Fh}

\maketitle

\newcommand{\ee}{\end{equation}}
\newcommand{\ba}{\begin{array}}
\newcommand{\ea}{\end{array}}
\newcommand{\beqn}{\begin{eqnarray}}
\newcommand{\eeqn}{\end{eqnarray}}

One of the most striking aspects of the human genome is the presence
of long stretches of DNA with no apparent (or known) significance
\cite{albe02_sh}. This is what biologists refer to as {\it junk} DNA,
and it comprizes the majority of our DNA. In the human genome (and
that of other higher eukaryotes) not only are the genes very sparse,
but most of them are interrupted by sequences, the introns, which are
non-coding i.e. they do not carry information for protein synthesis
\cite{albe02_sh}. During transcription introns are therefore removed
from the messenger RNA (mRNA), which is assembled only from the expressed
parts of the gene, the exons. In the human genome introns are on average
ten times longer than exons and thus constitute the majority of the
gene. Procaryotes (such as bacteria) instead have a very compact genome
without introns \cite{albe02_sh}.

The discovery of introns in 1977 triggered a debate around their
significance and origin, which lead to the formulation of the ``introns
early" \cite{gilb78,blak78,gilbert} and the ``introns-late" theories
\cite{cava85,dibb89,stol94_sh}. According to the ``introns early"
viewpoint the introns appeared at the origin of life and the exons were
small ancient genes. The bacteria then lost the introns due to selective
pressure in order to keep their genome short. The ``introns late"-theory
instead claims that introns must have appeared much later, ie during the
early eukaryotic evolution. A consensus between these opposing views has
meanwhile been reached in recent years. The analysis of an increasing
number of genes showed that most of the introns have a ``recent" origin,
although few are still believed to be very old \cite{roy02}. The mechanism
by which introns were included into the genome is, however, still poorly
understood (for a recent discussion, see eg. Ref \cite{cogh04}).

\begin{figure}[b]
\includegraphics[width=7.5cm]{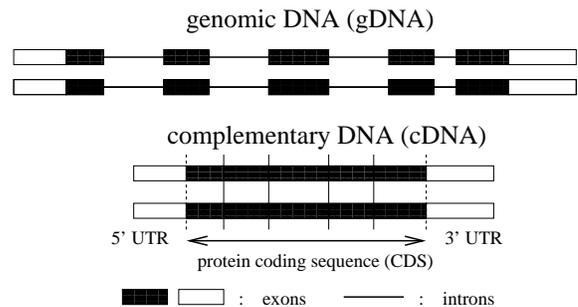}
\vspace{-2mm}
\caption{Schematic view of genomic and cDNA. {\it Exons} are the segments
of the gene transcribed into mRNA, while {\it introns} are spliced
out. The cDNA, which is a reverse-transcription from the single stranded
mRNA, contains no introns. It is characterized by exon-exon boundaries
(vertical solid lines) and boundaries between the protein coding sequence
(CDS) and untranslated regions (UTR) (vertical dashed lines).  The part of
the exons shown in black contains the protein coding sequence.  The $3'$
and $5'$ ends refer to those of the single stranded mRNA.}
\label{FIG01}
\end{figure}

In this paper we present the results of a study of the physical properties of
human DNA sequences which points to a possible pathway leading to intron
insertion in genes. By means of a statistical mechanics approach we
analyze the thermodynamic stability (``melting") of DNA sequences obtained
by assembling the exons together. This is known as complementary DNA
(cDNA) and can be obtained in the laboratory by reverse transcription of
mRNA. As illustrated in Fig. \ref{FIG01}, cDNA is characterized by
exon-exon boundaries and the boundaries between the coding sequence (CDS)
and the untranslated region (UTR). If introns were inserted ``recently"
into the genome, then the cDNA roughly resembles an ancient gene, apart
from the mutations which occurred since the insertion of the first introns
(see below). We find that exon-exon boundaries in cDNA sequences are
strongly correlated with their melting domains.

DNA melting is the process by which the double-stranded molecule
in solution dissociates into two separate strands by an increase of
temperature \cite{wart85}.  Fragments which are longer than $1000$ base
pairs dissociate through a multistep process in which different parts
of the chain melt at different temperatures. These ``melting domains"
are typically a few hundreds of nucleotides long.
The thermodynamics of the DNA melting process has been investigated both
experimentally \cite{wart85} and by means of numerical calculations
based on the statistical mechanics of the dissociation process
\cite{pola70,pola74}. The latter approach allows one to calculate
$\theta_i$, the probability that the $i$-th base pair is bound at a
temperature $T$. The total fraction of bound base pairs is then $\theta
= \sum_i \theta_i/N$, where $N$ is the number of nucleotide pairs in the
molecule.  The multistep nature of the melting transition can be seen in
a plot of $\theta$ vs. $T$. This quantity vanishes while increasing the
temperature through a series of jumps, which correspond to sharp peaks
in a plot of $-d \theta/dT$ versus $T$, the differential melting curve.
The parameter $\theta$ can also be measured by UV absorption of DNA
in solution \cite{wart85}. Typically statistical mechanics programs
reproduce the experimental results quite well \cite{blak99_sh}.

\begin{figure}[t]
\includegraphics[width=7.5cm]{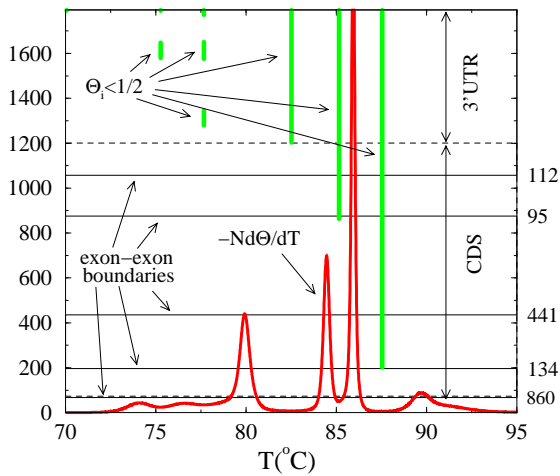}
\vspace{-2mm}
\caption{Differential melting curve and melting domains for the
human $\beta$-actin cDNA (NCBI entry code: NM\_001101). Horizontal axis:
temperature, vertical: $-N d\theta/dT$, and sequence length. Vertical bars
in the graph indicate the regions along the chain for which $\theta_i
< 1/2$.  Horizontal solid lines are exon-exon boundaries and dashed
lines are boundaries between the protein coding part of the sequence
(CDS) and the untranslated regions (UTR). A remarkable overlap between
the genomic and thermodynamic domains is observed.}
\label{FIG02}
\end{figure}

\begin{figure}[b]
\includegraphics[width=7.5cm]{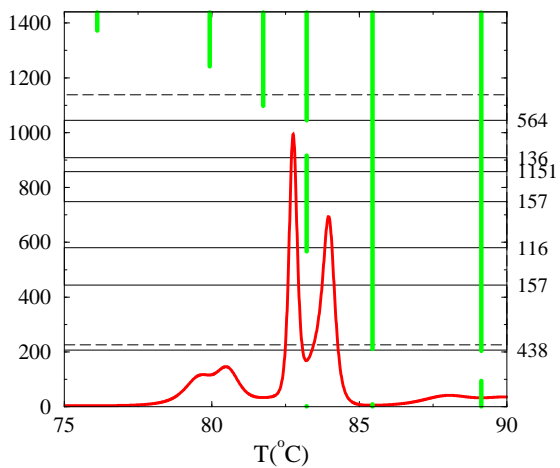}
\vspace{-2mm}
\caption{As in Fig. \ref{FIG02}, but 
for the human CDK4 cDNA (NCBI entry NM\_000075).}
\label{FIG03}
\end{figure}

In Fig. \ref{FIG02} we plot the melting curve $-N d \theta/dT$, as obtained
by a statistical mechanical calculation \cite{blak99_sh}, for the
human $\beta$-actin cDNA, where $N$ is the total length of the sequence
($N=1792$ in this case).  We used the same stacking energies and loop
entropic parameters as in Ref. \cite{blos03}. The salt concentration was
fixed at $0.05$ M. The three main melting peaks of Fig. \ref{FIG02}
indicate three sharp subtransitions which characterize the melting
of the sequence. The evolution of the average configuration of the
sequence as a function of $T$ can be read-off from the vertical bars,
which denote, at the given temperatures, the regions which are more
likely to be dissociated, i.e. where $\theta_i < 1/2$. For instance,
the bar shown at $T \approx 85^\circ C$ indicates that the region with
$i \gtrsim 850$ is dissociated, while that with $i \lesssim 850$ is in a
helical state. These melting domains are plotted for temperatures between
melting peaks, so that, by comparing the configurations at temperatures
below and above each peak, one can visualize the regions of the sequence
involved in the multistep melting.  We refer to the nucleotides separating
two neighboring regions of the sequence with $\theta < 1/2$ and $\theta >
1/2$ as the {\it thermodynamic boundaries}.  In Fig. \ref{FIG02} exon-exon
boundaries are indicated as horizontal solid lines, while the boundaries
between the CDS and UTRs are shown as dashed lines.  The numbers on the
left vertical axis, located at the exon-exon boundaries, show the length
on the introns in the genomic DNA.

\begin{figure}[t]
\includegraphics[width=7.5cm]{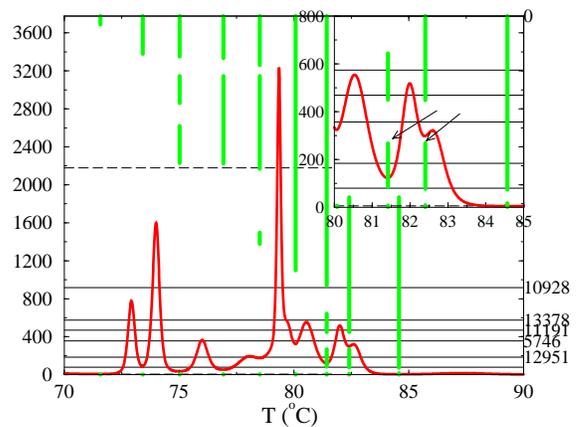}
\vspace{-2mm}
\caption{As in Fig. \ref{FIG02}, but for the EHHADH human gene (NCBI
entry NM\_001966).}
\label{FIG04}
\end{figure}

In the example of Fig. \ref{FIG02} the melting process starts with
the opening of small loops in the 3'UTR, while the first sharp peak at
$80^\circ C$ is the dissociation of the whole 3'UTR region. The next
peak at about $84^\circ C$ is due to the melting of the exons 5 and 6
(numbering them from the 5' region), while the melting of the exons 3 and
4 occurs at higher temperature ($\approx 86 ^\circ C$).  A remarkable
overlap between the locations of the thermodynamic and genomic domains
is observed.

An equally striking correspondence is found in most of the human cDNA
sequences we investigated. Figure \ref{FIG03} shows the melting curve for
the cDNA of the Cyclin dependent kinase (CDK4).  Occasionally, we found
some discrepancies, as can be seen e.g. in Fig. \ref{FIG04} which shows
the cDNA for the human HAADH gene.  The inset shows an enlargement of
the region for the temperature interval of $80$-$85^\circ C$.  Note that
the thermodynamic boundary indicated by the arrows splits the third exon 
of the sequence in roughly two equal parts.

Long cDNA sequences ($\gtrsim 3000$ bp) may have a very complex melting
curve with many overlapping peaks. In order to have a better criterion
for the definition of thermodynamic boundaries we have performed a
temperature scan from $60$ to $100^\circ C$ and calculated the boundaries
separating the $\theta_i < 1/2$ to the $\theta_i > 1/2$ regions at
a fixed interval $\Delta T = 0.01^\circ C$.  We have then derived a
histogram $h_i$ over all base pairs $i$ as follows: if $i$ is found to
be a thermodynamic boundary between two temperatures $T_1$ and $T_2 >
T_1$, the contribution to the histogram is $h_i = (T_2 - T_1)/\Delta T$.

\begin{figure}[t]
\includegraphics[width=8.5cm]{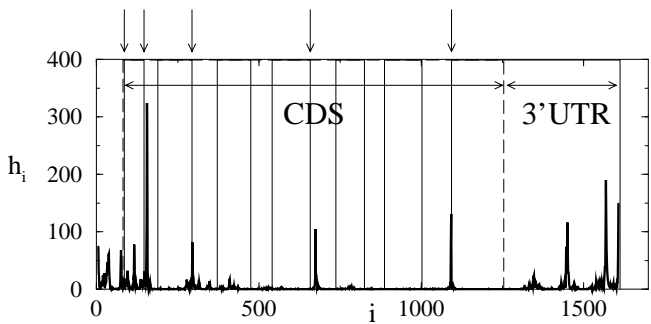}
\vspace{-2mm}
\caption{Histogram of the thermodynamic boundaries (think line) of the
Human cDNA encoding for the interleukin enhancer binding factor 2 (ILF2)
protein with Gene Bank entry NM\_004515.  The thin vertical lines denote
exon-exon (solid) and CDS-UTR (dashed) boundaries.  The arrows indicate
the exon-exon boundaries detected by the melting analysis.}
\label{FIG05}
\end{figure}

Figure \ref{FIG05} shows the histogram $h_i$ as function of $i$ for
the Human ILF2 cDNA (thick line). The solid and dashed thin vertical
lines denote the exon-exon and CDS-UTR boundaries, respectively. The
histogram is characterized by few main thermodynamic boundaries,
well above the noise level (as typically observed in all cases). The
advantage of $h_i$ with respect to a plot of the differential melting
curves is that boundaries appear more clearly also for long sequences,
and their stabilities can be quantified from the height of $h_i$. The
ILF2 cDNA melting of Fig. \ref{FIG05} is yet another example of the good
correspondence between thermodynamics and genomic features.  The exon-exon
boundaries ``detected" by thermodynamics are shown as vertical arrows
in Fig. \ref{FIG05}.

\begin{figure}[t]
\includegraphics[width=8.0cm]{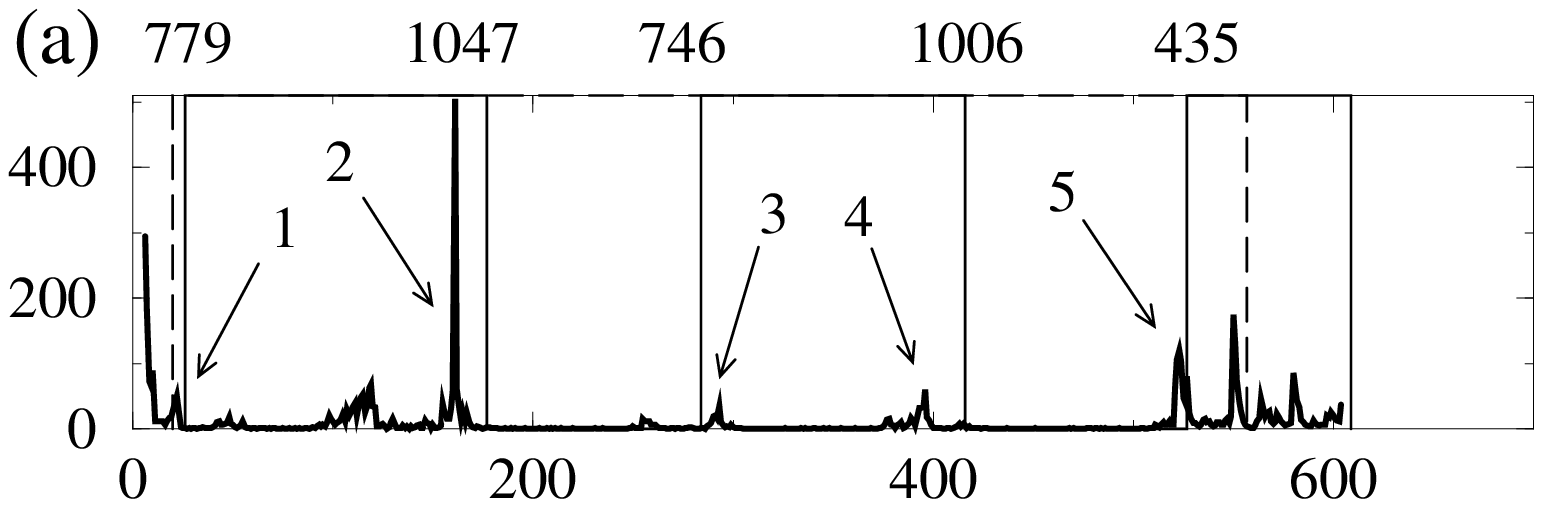}
\includegraphics[width=8.0cm]{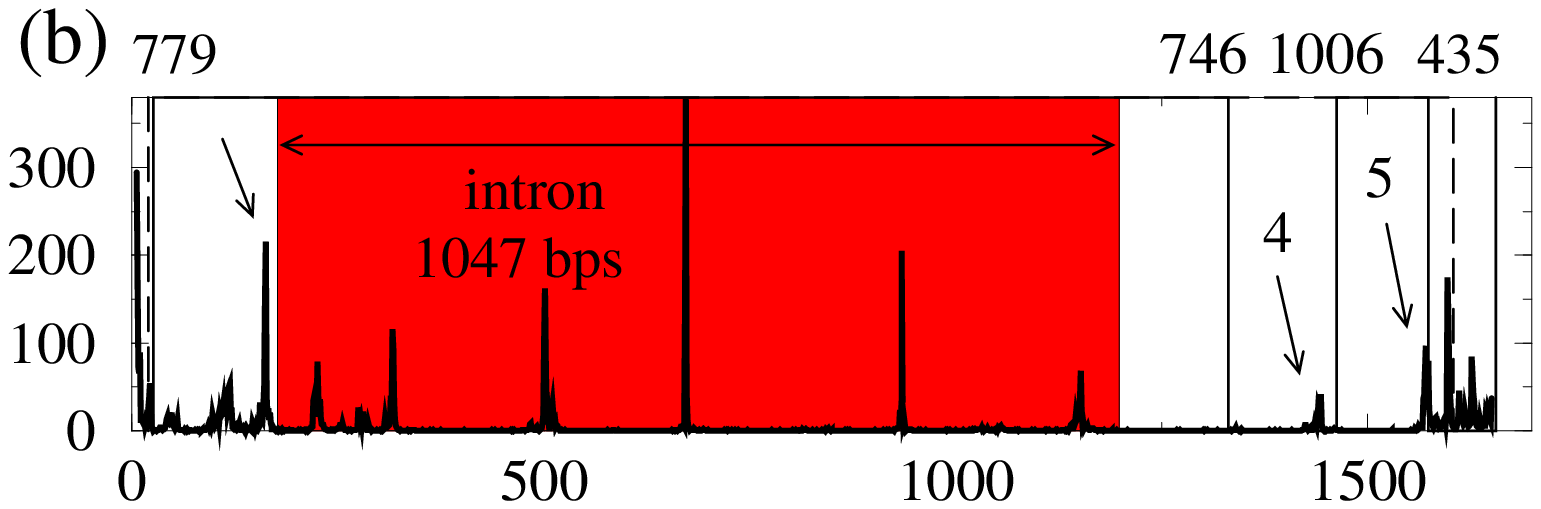}
\vspace{-2mm}
\caption{(a) Melting histogram for the cDNA sequence encoding for the 
ribosomal protein L11 (RPL11) with NCBI entry NM\_000975. The numbers
above the exon-exon boundaries denote the length of introns in the 
genomic DNA. (b) The same sequence as in (a) to which the intron 2
of 1047 bp is added (in color).}
\label{FIG06}
\end{figure}

The correspondence between melting domains and genomic features has
also been explored recently in studies of lower eukaryotes as {\it
S. Cerevesiae} \cite{yera00}, {\it P. Falciparum} \cite{yera00b} or {\it
D. Discoideum} \cite{marx98_sh}. These studies focused on {\it genomic}
DNA and in particular on the correspondence between exon-intron and
thermodynamic boundaries. This correspondence allowed Yeramian {\it
et al.} to locate genes in the {\it P. Falciparum} \cite{yera02_sh}
and {\it D. Melanoganster} \cite{yera03} genome. Exon-intron boundaries
may be difficult to detect in higher eukaryotes by melting analysis as
introns tend to be very long. We have illustrated this in Fig. \ref{FIG06}
which shows the melting histogram for the human ribosomal protein L11
cDNA sequence (a) and for the cDNA in which the second intron has been
inserted.  In the cDNA the strongest peaks in the histogram are correlated
with exon-exon boundaries this holds for the boundaries 1,2 and 5 (notice
the thermodynamic boundary 2 is shifted of 15 base pairs compared to the
exon-exon boundary), but notice also some weaker signals close to the
boundaries 3 and 4 (such weak signals have not been taken into account
in the histogram of Fig. \ref{FIG06}). When the intron is inserted
(Fig. \ref{FIG06}(b)) many other ``spurious" peaks appear, therefore
exon-introns boundaries may be difficult to detect from thermodynamics,
although they still persist.

\begin{figure}[t]
\includegraphics[height=4.8cm]{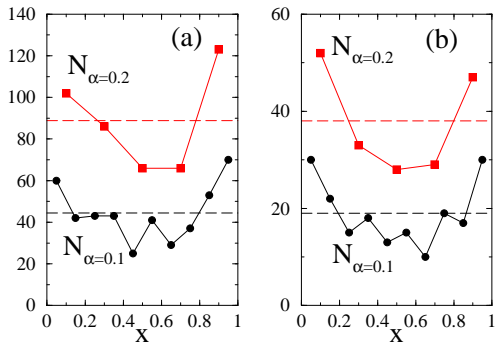}
\vspace{-2mm}
\caption{Plots of $N_\alpha (x)$, the number of observed thermodynamic
boundaries in the interval $x-\alpha/2$, $x+\alpha/2$. (a) Set of $48$
human cDNAs selected at random from the Refseq Genbank set \cite{ncbi}
and (b) set of $35$ human cDNA from Housekeeping genes taken from
\cite{HK}. The dashed lines corresponds to a random distribution of
uncorrelated exon-exon and thermodynamic boundaries.}
\label{FIG07}
\end{figure}

We have analyzed $48$ genes for which exons-introns boundaries have been
experimentally confirmed, selected randomly from the GenBank Refseq set
\cite{ncbi}, and $35$ housekeeping (HK) genes, taken from Ref. \cite{HK}
(HK genes are virtually expressed in all tissues \cite{albe02_sh}). For
each sequence we have produced a histogram of thermodynamic boundaries
as those of Figs. \ref{FIG05} and \ref{FIG06}. We recorded the position
of the major peaks and calculated the scaled distance from exon-exon
boundaries. As an example, if a thermodynamic boundary is found at
position $i_{\rm th}$ and it is contained in an exon beginning at
$i_1$ and ending at $i_2$ the scaled distance is defined as $x=(i_{\rm
th}-i_1)/(i_2-i_1)$, thus $0 \leq x \leq 1$.  Figure \ref{FIG07} shows a
plot of the function $N_\alpha (x)$, defined as the number of observed
thermodynamic boundaries in the interval $x-\alpha/2$, $x+\alpha/2$.
For reference, this result is compared to that obtained from random
distributions of uncorrelated boundaries, which is a constant
(dashed lines in Fig. \ref{FIG07}). The plot clearly demonstrates
the significance of the observed correlations.  As is clear from the
Figs. \ref{FIG02}-\ref{FIG05} not all exon-exon boundaries are "detected"
by thermodynamics. Our statistical analysis indicates that the detection
score is of roughly $35\%$.

\begin{figure}[t]
\includegraphics[width=6.7cm]{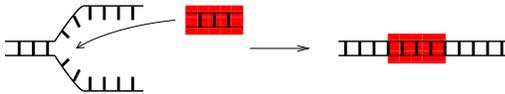}
\vspace{-2mm}
\caption{Schematic view of a possible pathway of intron insertion in
the genome: introns seem to have targeted preferentially thermodynamic
boundaries.}
\label{FIG08}
\end{figure}

In our view, the correspondence between exon-exon and thermodynamic
boundaries suggests a possible pathway of introns insertion in the
genome during evolution. Introns seem to have targeted exposed bases at a
``fork" between a double helix and an open DNA region as schematically
shown in Fig. \ref{FIG08}. The exposed bases could have provided sites
where binding with ``foreign" introns sequences was possible, probably
because the two strands are still sufficiently close to each other to
integrate efficiently the intron. To our knowledge, the idea that the
thermodynamic stability of the double stranded DNA played a role in
the intron insertion in eukaryotic genomes has not been considered in
other studies.  The emphasis so far has been put on the specific sequence
composition of few base pairs around the insertion site; for instance
it has been shown \cite{dibb89} that the upstream exon tends to end
with (C/A)AG and the downstream exon tends to start with GT, which were
interpreted as sequence specific targeting of some still uncharacterized
intron insertion machinery.  As the precise biochemical mechanism for
the introns insertion has not yet been understood, both hypothesis of
insertion at specific sequence sites or at thermodynamic boundaries remain
plausible.  Since about 1/3 of the exon-exon boundaries are ``detected"
by our thermodynamic analysis, it is also possible that other mechanisms
of introns insertion may have been used during evolution as well.
A systematic study of the relationship between thermal and exon-exon
boundaries combined with a phylogenetic analysis for different species
in the eukaryotic kingdom, may provide further insights in these issues.

Finally, the boundaries separating regions of DNA with different stability
properties, found in the melting analysis, should manifest themselves
also under non-equilibrium conditions.  Mechanical unzipping of DNA,
for instance, is known to generate metastable forks \cite{week04_sh}.
Moreover the thermodynamic boundaries obtained from statistical mechanics
approaches are robust with respect to ``realistic" changes in the stacking
energies and entropic parameters \cite{yera00,blos03}. These modifications
do not affect the location of the thermodynamic boundaries noticeably.
Likewise, as we explicitly verified, a small percentage of mutations
does not have a strong influence on the thermodynamic boundaries.
As the coding parts of the genes tend to be highly conserved across
distant species \cite{albe02_sh}, we expect that the melting features
of cDNA sequences are very similar to that of the old ``protogenes"
as they were before the intron insertions.

%

 \end{document}